\documentclass[%
 reprint,
superscriptaddress,
footinbib,
 amsmath,amssymb,
 aps,
pra,
floatfix,
]{revtex4-1}

\usepackage{tabularx}
\usepackage{graphicx}
\usepackage{dcolumn}
\usepackage{bm}

\begin{document}

\title{Gate-defined, Accumulation-mode Quantum Dots in Monolayer and Bilayer WSe$_2$}

\author{S.~Davari}
\affiliation{Department of Physics, University of Arkansas, Fayetteville, AR 72701}
\author{J.~Stacy} 
\affiliation{Department of Physics, University of Arkansas, Fayetteville, AR 72701}
\author{A.~M. Mercado}
\affiliation{Department of Physics, University of Arkansas, Fayetteville, AR 72701}
\author{J.~D.~Tull}
\affiliation{Department of Physics, University of Arkansas, Fayetteville, AR 72701}
\author{R.~Basnet} 
\affiliation{Department of Physics, University of Arkansas, Fayetteville, AR 72701}
\author{K.~Pandey}
\affiliation{Department of Physics, University of Arkansas, Fayetteville, AR 72701}
\author{K.~Watanabe}
\affiliation{Advanced Materials Laboratory, National Institute for Materials Science, 1-1 Namiki, Tsukuba, Ibaraki 305-0044, Japan}
\author{T.~Taniguchi} 
\affiliation{Advanced Materials Laboratory, National Institute for Materials Science, 1-1 Namiki, Tsukuba, Ibaraki 305-0044, Japan}
\author{J.~Hu}
\affiliation{Department of Physics, University of Arkansas, Fayetteville, AR 72701}
\author{H.~O.~H.~Churchill}
\email{churchill@uark.edu}
\affiliation{Department of Physics, University of Arkansas, Fayetteville, AR 72701}

\begin{abstract}
We report the fabrication and characterization of gate-defined hole quantum dots in monolayer and bilayer WSe$_2$. The devices were operated with gates above and below the WSe$_2$ layer to accumulate a hole gas, which for some devices was then selectively depleted to define the dot. Temperature dependence of conductance in the Coulomb blockade regime is consistent with transport through a single level, and excited state transport through the dots was observed at temperatures up to 10 K. For adjacent charge states of a bilayer WSe$_2$ dot, magnetic field dependence of excited state energies was used to estimate $g$-factors between 0.8 and 2.4 for different states.  These devices provide a platform to evaluate valley-spin states in monolayer and bilayer WSe$_2$ for application as qubits.
\end{abstract}
\date{\today}
\maketitle

\section{\label{sec:intro}Introduction}

Certain crystals possess two or more inequivalent band extrema, or valleys, that can serve as a pseudospin defining a qubit.
Application of the valley and spin degrees of freedom for qubits was first demonstrated in carbon nanotube quantum dots \cite{laird2013valley} and has also been investigated in Si quantum dots \cite{mi2018landau,penthorn2019two}. 
As an alternative, coherent valleytronics using few-layer transition metal dichalcogenides (TMDs) offers long-lived and coherent valley-spin states \cite{xu2014spin,hao2016direct,jones2013optical,tang2019long,lu2019optical,yang2015long,song2016long,dey2017gate,kim2017observation,jin2018imaging,rivera2016valley}, several proposals for qubit designs \cite{gong2013magnetoelectric,wu2016spin,brooks2017spin,szechenyi2018impurity,pawlowski2018valley,david2018effective,pawlowski2019spin,chen2020theory}, and an inherent light-matter interface for control, readout, and coupling to other quantum systems \cite{mak2016photonics}.

Confinement sufficient to address valley-spin states of a small number of particles is required to define a qubit in this context, and significant recent activity in this area has focused on quantum emitters localized by defects \cite{chakraborty2015voltage,koperski2015single,he2015single,srivastava2015optically}, strain \cite{kumar2015strain,palacios2017large,branny2017deterministic}, and moir\'{e} patterns \cite{tran2019evidence,seyler2019signatures,alexeev2019resonantly,jin2019observation}.
For the valley pseudospin, localization to length scales not significantly larger than the lattice spacing generates valley coupling that would be detrimental to the design and coherence of an eventual TMD valley-spin qubit \cite{liu2014intervalley}.
Gate-defined quantum dots, in addition to allowing electronic probes and electrical control of devices, provide a means to tune the confinement length $L$ in valley-based qubits to balance the competing demands of valley coupling in small dots against impractically small level spacing $\Delta\propto1/L^2$ in large dots.

TMD single quantum dots defined by gates have been reported for multi-layer WSe$_2$ and WS$_2$ \cite{song2015gate, song2015temperature} and few-layer MoS$_2$ \cite{lee2016coulomb,wang2018electrical, pisoni2018gate}, and a double quantum dot has been reported in multilayer MoS$_2$ \cite{zhang2017electrotunable}. 
Additionally, excited states have been observed in a dot defined in a MoS$_2$ nanotube \cite{reinhardt2019coulomb}, and size-controlled dots have also been formed in etched MoS$_2$ \cite{wei2017size}.
All two-dimensional TMD quantum dot devices reported so far have operated in a so-called classical limit, $\Delta< k_B T$, in which the individual quantum states required to define an eventual qubit were not resolved \cite{kouwenhoven1997electron}.
Additionally, there have been no reports of gate-defined quantum dots using monolayer and bilayer WSe$_2$, which has so far been the most actively investigated material for non-gate-defined quantum emitters.

In this Letter we report the fabrication and characterization of gate-defined hole quantum dots using monolayer and bilayer WSe$_2$ that have $L\sim25$ nm, small enough for the observation of discrete levels at temperatures up to 10 K.
Because of the low level of unintentional doping in monolayer and bilayer WSe$_2$, an additional accumulation gate was required to generate a hole gas within which the dot could be defined, similar to an approach that has been used successfully for quantum dots in Si \cite{zwanenburg2013silicon,kawakami2014electrical}.
Temperature dependence of a Coulomb blockade conductance peak maximum confirms single-level transport, and comparison of excited state energies at different magnetic fields provides a lower-bound on the $g$-factors of excited states of adjacent charge states in a bilayer WSe$_2$ quantum dot ranging from $g=0.8$ to 2.4.

\begin{figure*}
\includegraphics[width=6.74in]{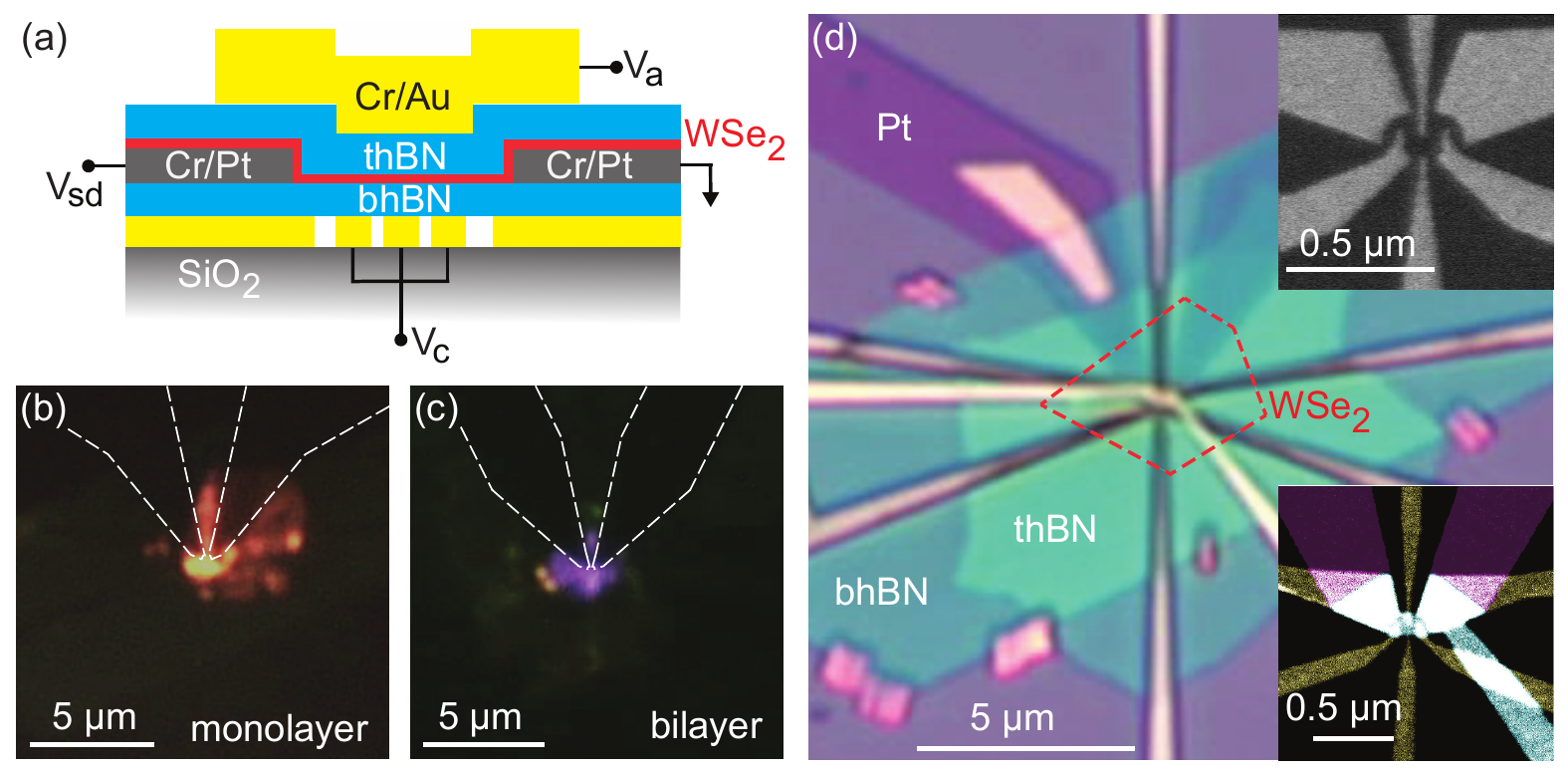}
\caption{\label{figure1}(a) Schematic of monolayer or bilayer WSe$_2$ quantum dot devices, contacted on bottom by Pt, encapsulated by top (thBN) and bottom (bhBN) layers, and with bottom confining gates and a top accumulation gate. (b) Photoluminescence images of a monolayer and (c) bilayer WSe$_2$ device with contacts outlined in white. (d) Optical image of a complete device (location of the WSe$_2$ outlined with red dashed line). Upper inset: scanning electron micrograph of the confining gates.  Lower inset: false-color scanning electron micrograph of a complete device showing alignment of the confining gates (yellow), contacts (purple), and accumulation gate (green).}
\end{figure*}

\section{\label{sec:expt}Experimental Methods}
Because of the relatively large band gap and heavy effective mass in WSe$_2$ compared to most other materials used for gate-defined quantum dots, two principles guided the design of our devices:  first, gate dimensions were made as small as possible to maximize $\Delta$, and second, metal contacts were brought as close as possible to the entrance and exit of the dot to avoid the creation of multiple accidental dots in the contact region. 

A schematic of the WSe$_2$ quantum dot devices we fabricated is shown in Fig.~1(a), in which monolayer or bilayer WSe$_2$ was encapsulated by two hBN layers and contacted from below with Pt.
The devices were gated on top with a single accumulation gate and on bottom with four confining gates.
In detail, fabrication began with lithography and evaporation (5/15 nm Cr/Au) of bottom confining gates defining a quantum dot with a lithographic diameter of $\sim$80-100 nm [upper right inset to Fig.~1(d)]. 
We note that the large gates under the contact region of the device were not used electrically, but were present to promote flatness of the final hBN/WSe$_2$/hBN stack (without these features, the bottom hBN layer invariably wrinkled in the dot region).
Next, the confining gates were insulated by polycarbonate-based dry transfer of the bottom hBN layer ($\sim$10 nm thick), which was tacked in place around the perimeter [Fig.~1(d)] to prevent lateral movement during the second transfer. 
Bottom contacts for the WSe$_2$ were then patterned on the bottom hBN layer using 2/8 nm of Cr/Pt deposited by electron beam evaporation \cite{Movva}. 
To remove processing residue from the bottom hBN, the devices were then annealed for 3 hours at 200 $^{\circ}$C in forming gas (3\% H$_2$ in Ar). 

Bulk WSe$_2$ grown by chemical vapor transport was mechanically exfoliated to obtain monolayer and bilayer flakes, which were identified and confirmed using photoluminescence imaging and spectroscopy \cite{zhao2013evolution}. 
hBN flakes were also obtained by exfoliation of bulk crystals, and the top hBN layer ($\sim$10 nm thick) and monolayer or bilayer WSe$_2$ were then picked up and transferred onto the contacts and bottom hBN.
Photoluminescence images for monolayer and bilayer WSe$_2$ at this stage of the fabrication process are shown in Figs.~1(b) and 1(c), respectively.
The top accumulation gate and bond pads (5 nm Cr and 40-65 nm Au) were then defined in a final lithography step. 
The accumulation gate is shaped such that holes accumulate only in the dot and contact regions of the WSe$_2$ flake.
The overlay of all three metal layers is shown in the lower right inset to Fig.~1(d) \footnote{To promote adhesion the accumulation gate was extended in later designs across the entire width of the top and bottom hBN layers, but this feature is not present in the lower right inset to Fig.~1(d).}.  
Finally, a second forming gas anneal identical to the first one was necessary to achieve adequate contact transparency for low-temperature, low-frequency transport measurements. 
An optical image of a completed device is shown in Fig.~1(d) with the location of the monolayer WSe$_2$ outlined in red.  
Devices were measured using standard DC transport and lock-in techniques in a pumped $^4$He cryostat (2 K) or a dilution refrigerator (50 mK).

\section{\label{sec:results}Results and Discussion}
Six devices (two monolayers and four bilayers) were measured, and we focus on electronic transport data for three of them, one monolayer device (denoted ML) and two bilayer devices (BL1 and BL2).
First, DC current at fixed source-drain bias $V_{\rm sd}$ was measured as a function of the voltage applied to the accumulation gate, $V_{\rm a}$.  
Both monolayer devices displayed ambipolar transport characteristics as shown in Fig.~2(a) for device ML at a temperature of 55 mK and $V_{\rm sd} = 12$ mV. 
Current strongly favored hole conduction, as expected for the high work function Pt contacts \cite{Movva}.
Bilayer devices were similar but with generally lower resistance, and in contrast to the monolayer devices, no measurable n-type current reliably distinct from leakage current was observed, even at a relatively high $V_{\rm sd} = 0.3$ V and higher temperature (2.3 K), as shown in Fig.~2(b).
\begin{figure}
\includegraphics[width=3.37in]{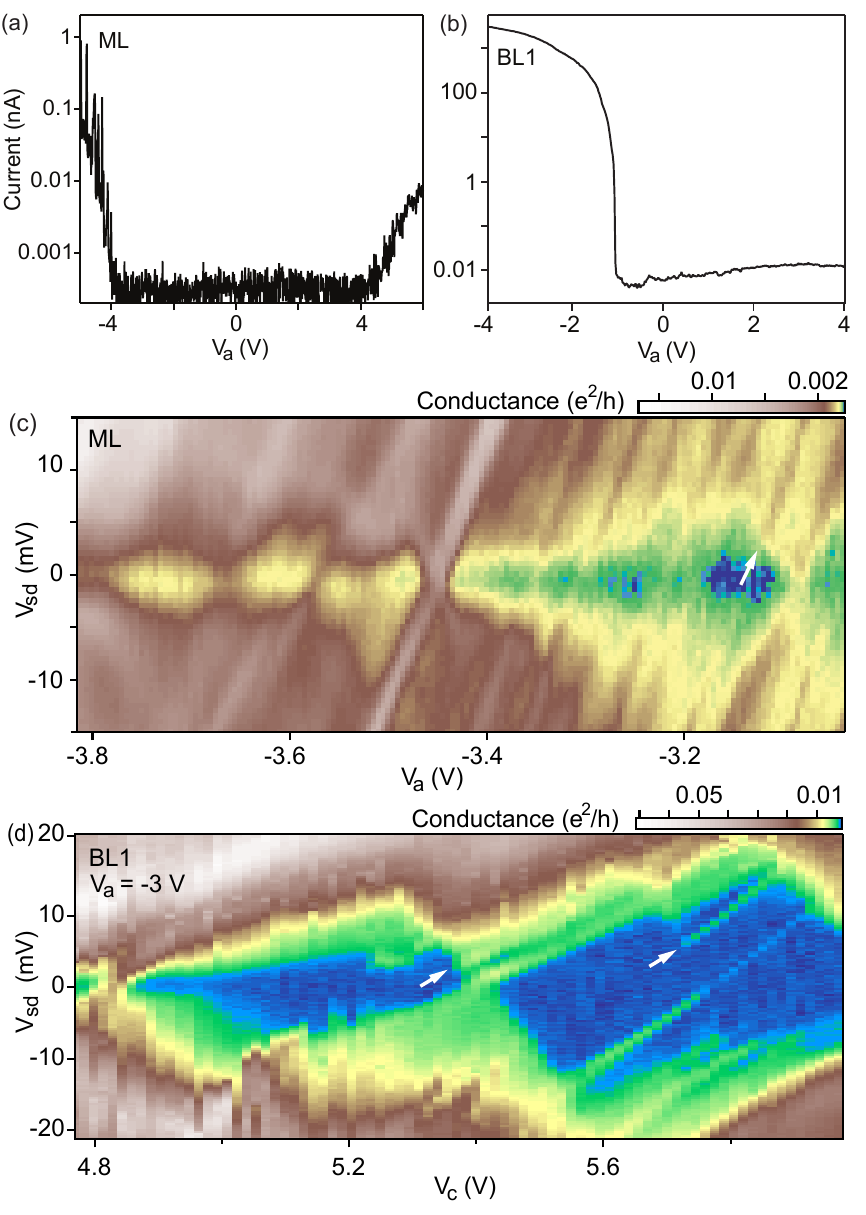}
\caption{\label{figure2}(a) Current as a function of $V_{\rm a}$ for monolayer device ML at a temperature of 50 mK. (b) Current as a function of $V_{\rm a}$ for bilayer device BL1 at a temperature of 2.1 K. (c) Conductance as a function of $V_{\rm sd}$ and $V_{\rm a}$ for device ML. (d) Conductance as a function of $V_{\rm sd}$ and $V_{\rm c}$ with $V_{\rm a} =$  -3 V.}
\end{figure}

In measurements of differential conductance as a function of $V_{\rm sd}$ and gate voltage ($V_{\rm a}$, confining gate voltage $V_{\rm c}$, or both), all six devices showed diamond-shaped conductance resonances indicative of Coulomb blockade.
Devices with higher resistance formed quantum dots when only $V_{\rm a}$ was swept (accumulation mode), presumably because of sufficiently high contact resistance to form tunnel barriers at the contacts.
One example of accumulation mode operation is shown in Fig.~2(c) for device ML, in which we note a transition from well-defined Coulomb diamonds around $V_{\rm a} = $ -3.1 V to a more open regime below $V_{\rm a} = $ -3.6 V.
Upward- and downward-moving features in Fig.~2(c) have slopes with nearly equal magnitude.
Defining contributions to the dot capacitance from the source, drain, accumulation gate, and all confining gates as $C_s$, $C_d$, $C_a$, and $C_c$, respectively, we expect features with slope magnitudes of $C_a/C_s$ and $C_a/(C_d+C_c)$ \cite{hanson2007spins}.
Because $C_d\gg C_c$, near-equality of slope magnitudes implies $C_s\simeq C_d$ and locates the dot approximately equidistant from each lead.

In lower resistance devices such as BL1 [Fig.~2(b)], activation of the confining gates was required to form a dot by depleting charge accumulated by $V_{\rm a}$.
Coulomb diamonds for this device as a function of $V_{\rm sd}$ and $V_{\rm c}$ at $V_{\rm a}=$ -3 V are shown in Fig.~2(d).
Here $V_{\rm c}$ is the voltage applied to all confining gates simultaneously.
In this case the Coulomb diamonds are tilted with unequal slopes for positively- and negatively-sloped features.
In a Coulomb diamond measured as a function of $V_{\rm c}$, the slopes are $C_c/C_s$ and $C_c/(C_d+C_a)$, and since $C_a\gg C_c$, diamonds as a function of $V_{\rm c}$ are expected to be tilted even for $C_s=C_d$.

We also note the appearance of clear finite-bias resonances in the Coulomb diamonds of both devices, marked with white arrows in Figs.~2(c) and (d).
For example, these resonances appear most prominently in Fig.~2(d) at positive bias for the charge transitions near $V_{\rm c}=5.4$ and $5.7$ V.
Such features may arise from excited valley, spin, and/or orbital states in the limit $\Delta > k_B T$, but various extrinsic mechanisms such as quasi-one-dimensional density of states fluctuations in the leads could also give rise to finite-bias conductance resonances even for $\Delta < k_BT$, particularly for devices such as ours with narrow leads \cite{schmidt1997spectroscopy,mottonen2010probe,escott2010resonant}.

The temperature dependence of the Coulomb blockade peak height is a standard method to distinguish between single- and multi-level transport through a quantum dot:
the peak resistance is linear in $T$ in the single-level case and independent of $T$ in the multi-level case \cite{beenakker1991theory, kouwenhoven1997electron}.
For example, Song {\it et al.}~demonstrated the multi-level, temperature-independent regime for a WS$_2$ quantum dot \cite{song2015temperature}.
In Fig.~3(a) we show the evolution of a Coulomb blockade peak at zero bias for temperatures from 2.1 to 10 K for a second bilayer WSe$_2$ device, BL2.
The resistance at the peak increases approximately linearly [Fig.~3(b)], consistent with single-level transport.
We therefore conclude that device BL2 and other lithographically identical devices are small enough that $\Delta > k_B T$, and we associate the finite-bias conductance resonances with excited states of the quantum dot.
Additionally, given our accumulation gate geometry in which $V_{\rm a}$ modulates the carrier density in the dot as well as the WSe$_2$ covering the contacts, we would not expect density of states resonances in the leads to be parallel to the Coulomb diamond edges.

\begin{figure}
\includegraphics[width=3.37in]{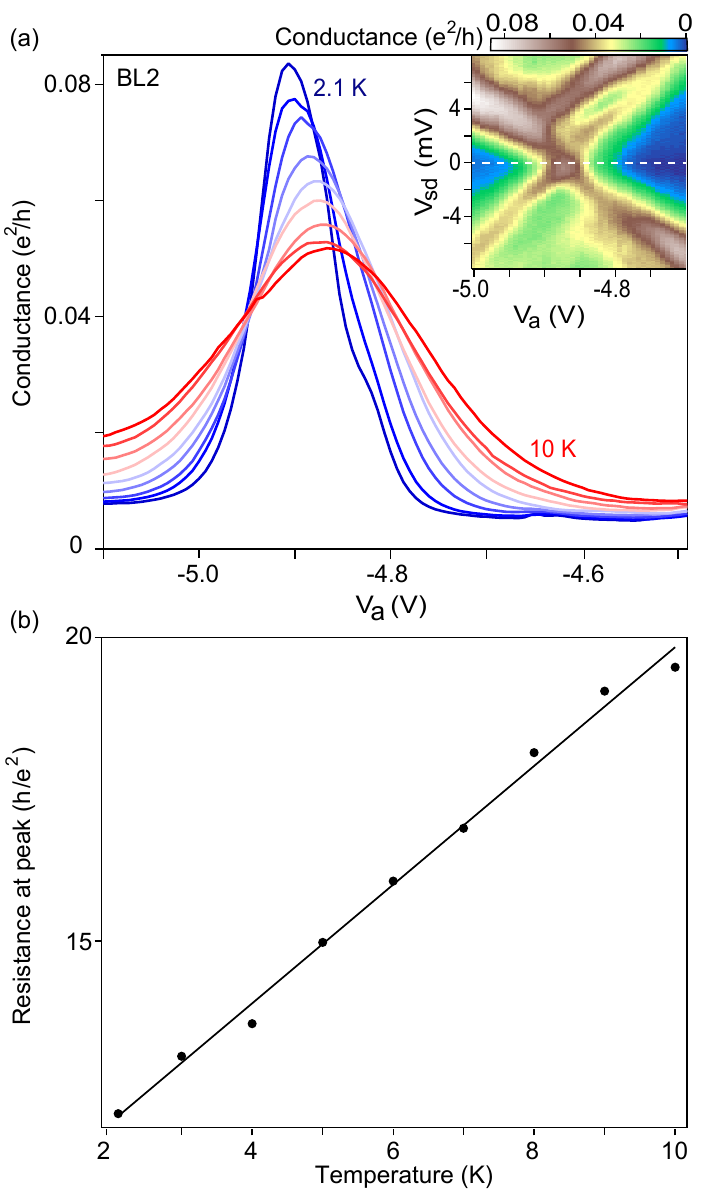}
\caption{\label{figure3}(a) Conductance at zero bias and zero magnetic field as a function of $V_{\rm a}$ at temperatures from 2.1 (blue) to 10 K (red) for device BL2.  Inset: Coulomb diamond associated with the data shown in (a), taken in a magnetic field of 7 T and at a temperature of 2.3 K. (b) Peak resistance of the curves in (a) as a function of temperature.  The solid black line is a linear fit.}
\end{figure}

Next we examine the magnetic field dependence of excited state energies for two adjacent charge states in device BL2 (Fig.~4).
We applied a field perpendicular to the WSe$_2$ because the large out-of-plane Zeeman-like spin splitting due to spin-orbit coupling \cite{zhu2011giant} suppresses in-plane moments of valley-spin states in monolayer TMDs \cite{zhang2017magnetic}.
This situation applies also to a bilayer TMD under an inversion symmetry-breaking transverse electric field \cite{wu2013electrical}, which is present in our devices in the range of 0.3-0.5 V/nm.

At zero field an excited state, denoted $\alpha$, is visible at 0.4 meV above the ground state of the charge state labeled $N$ in Fig.~4(a).
In a perpendicular magnetic field of 8 T, this state moves up to an energy of 0.8 meV without splitting, and for the $N$-$1$ charge state, an excited state, $\beta$, appears at 0.9 meV while the ground state conductance is suppressed [Fig.~4(b)].
Additionally, the state labeled $\gamma$ in Fig.~4(a) splits by 1.1 meV, as indicated by the dashed black lines in Fig.~4(b).
Based on these energy shifts of $E_Z = g\mu_B B$, we calculate $g$-factors of 0.8, 1.9, and 2.4 for $\alpha$, $\beta$, and $\gamma$, respectively.
These estimates represent lower bounds on the $g$-factors because this method of measurement is insensitive to intermediate level crossings that may occur between 0 and 8 T.

We emphasize that these $g$-factors, which are significantly smaller than those observed for confined WSe$_2$ excitons \cite{chakraborty2015voltage,he2015single,koperski2015single,srivastava2015optically}, are not expected be generic to all WSe$_2$ quantum dots, different even/odd charge state pairs within the same dot, or even different levels of the same charge state.
First, in addition to dot orbital effects in perpendicular field, the Zeeman energy in WSe$_2$ hole quantum dots has contributions from spin, valley, and atomic orbital magnetic moments \cite{xiao2012coupled} which in a simple model could individually combine to add or subtract from the total moment of a given state \cite{koperski2018orbital}, or that may each be coupled in complicated ways through device-specific parameters.
Second, while the dominant contribution to spin-orbit coupling in TMDs is Zeeman-like, Rashba spin-orbit coupling is expected to be non-negligible in aggressively gated devices \cite{kormanyos2014spin}, and $g$-factors may therefore be dependent on device geometry and electric fields as well \cite{bjork2005tunable,csonka2008giant}.

\begin{figure}
\includegraphics[width=3.37in]{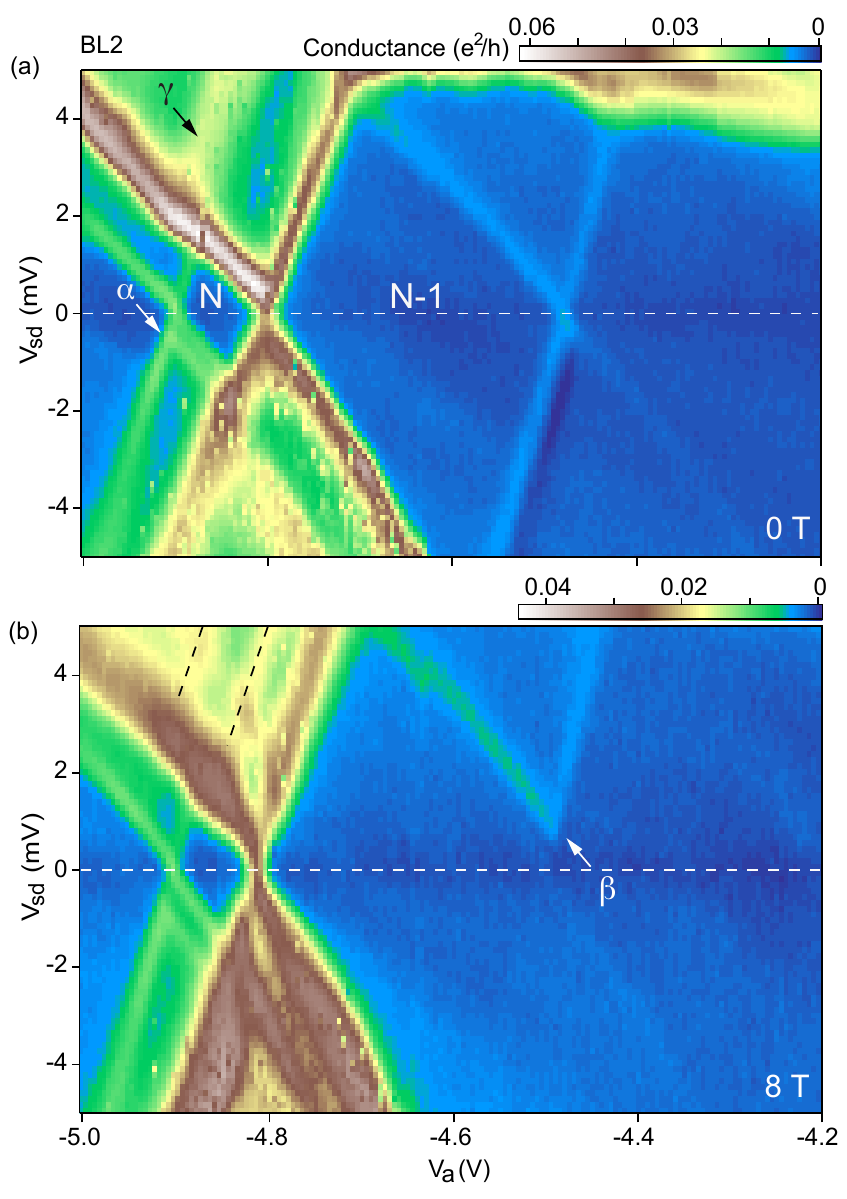}
\caption{\label{figure4}(a) Conductance as a function of $V_{\rm sd}$ and $V_{\rm a}$ at zero magnetic field and a temperature of 50 mK for device BL2. (b) Same as (a) in a perpendicular magnetic field of 8 T.}
\end{figure}

Given the expected importance of dot size to valley coupling strength and the performance of eventual coherent valleytronic devices, we roughly estimate the diameter of our dots using the gate capacitance.
The change in gate voltage required to add one hole to the dot is $$\Delta V_a = \frac{e}{C_a}\left(1+\frac{\Delta}{E_C}\right),$$
where $C_a$ is the accumulation gate capacitance and $E_C$ is the charging energy \cite{kouwenhoven1991single}.
For the charge state labeled $N$-$1$ in Fig.~4(b), the addition energy is $E_{\rm add} \equiv E_C + \Delta = 4.8$ meV.
To obtain $\Delta$, we note that the $(N$-$1)$-hole excited state $\gamma$ at $V_{rm a} = -4.9$ V in Fig.~4(a) moves with $V_{\rm a}$ at the same slope as the boundaries of the diamond and that $\gamma$ splits in a magnetic field.
These observations strongly suggest that $\gamma$ is an orbital excited state of the dot with $\Delta = 2$ meV \cite{escott2010resonant}.

Using these parameters and with $\Delta V_{\rm a} = 0.32$ V, we obtain $C_a\sim0.9$ aF.
The planar geometry of the device in which the accumulation gate covers the entire area of the dot permits us to model the accumulation gate and dot as a parallel plate capacitor with the relative permittivity of hBN assumed to be in the range of $2.5<\epsilon_{\rm hBN}< 3.5$ \cite{hunt2013massive,ahmed2018dielectric}.
The top hBN thickness for device BL2 was 7 nm, which yields a dot diameter $L\sim15$-20 nm depending on the value of $\epsilon_{\rm hBN}$.
A similar analysis for device BL1 yields a slightly larger dot size in the range 20-25 nm \footnote{Here we used the ${V}_a$-dependence of the charge state at ${V}_c=5.6$ {V} in {F}ig.~2(d)}. 
These sizes are a factor of 3-5 larger than defect- or strain-bound quantum dots in TMDs \cite{li2019dipolar} and are comparable to the size of the moir\'{e} supercell in moir\'{e}-defined dots \cite{tran2019evidence,seyler2019signatures,alexeev2019resonantly,jin2019observation}.
We also note that a particle-in-a-box estimate for $\Delta\sim \hbar^2/mL^2$ is on the order of 1 meV assuming a hole mass of $0.45m_0$ \cite{Movva}, which roughly agrees with the zero-field energy of state $\gamma$.

Finally, we estimate the number of holes in our dots.
Hole conduction turns on at room temperature in our devices at electric fields from the accumulation gate of approximately 0.2 V/nm, and a typical operating point for the dots at low temperature is 0.4 V/nm.
The difference in threshold and operating voltage yields a hole density $n\sim3\times10^{12}$ cm$^{-2}$, and multiplying by the dot area estimated above yields a hole number that could be expected to lie between 10 and 20.  
Here the uncertainty is dominated by the use of room-temperature threshold voltage to estimate the location in gate voltage of zero density.
\section{\label{sec:conclusion}Conclusion}
We have fabricated small monolayer and bilayer WSe$_2$ quantum dots using a combination of accumulation and confining gates.
Six devices showed Coulomb blockade at temperatures below 10 K, and three devices (one monolayer and two bilayers) showed conductance resonances consistent with single-level transport, as confirmed by temperature dependence of the peak height for one of the devices.
Magnetic field dependence of peak positions implied $g$-factors ranging from 0.8 to 2.4 in one bilayer device.

These observations constitute the first gate-defined quantum dots in monolayer and bilayer WSe$_2$ and the first observation of well-resolved excited states in two-dimensional TMD quantum dots.
In addition to satisfying essential prerequisites for the development of future electronic and/or optoelectronic qubits based on valley-spin states in few-layer TMDs, these devices also provide a platform for high-resolution, fundamental investigations of those states using Coulomb blockade spectroscopy and other techniques.

\begin{acknowledgments}
We acknowledge support from AFOSR award number FA9550-16-1-0203.  WSe$_2$ single crystal growth is supported by the US Department of Energy, Office of Science, Basic Energy Sciences program under award DE-SC0019467.
\end{acknowledgments}

\bibliographystyle{apsrev4-1}

%


\end{document}